\documentclass[10pt,conference]{IEEEtran}

\usepackage[utf8]{inputenc}
\usepackage{graphicx}
\usepackage{xcolor}
\usepackage{subcaption}
\usepackage{wrapfig}
\usepackage{hyperref}
\usepackage[inline]{enumitem}
\usepackage{footnote}
\usepackage{footmisc}
\usepackage{ifthen}
\usepackage{tcolorbox}
\usepackage{booktabs}
\usepackage{cite}
\usepackage{xspace}
\newcommand{\ie}{\emph{i.e.,}\xspace}
\newcommand{\eg}{\emph{e.g.,}\xspace}
\newcommand{\etc}{etc.\xspace}
\newcommand{\etal}{\emph{et~al.}\xspace}

\graphicspath{{figures/}}

\makeatletter
\def\ps@IEEEtitlepagestyle{
	\def\@oddfoot{\mycopyrightnotice}
	\def\@evenfoot{}
}
\def\mycopyrightnotice{
	{\footnotesize
		\begin{minipage}{\textwidth}
			\centering
			\textcopyright~2021 IEEE.  Personal use of this material is permitted.  Permission from IEEE must be obtained for all other uses, in any current or future media, including reprinting/republishing this material for advertising or promotional purposes, creating new collective works, for resale or redistribution to servers or lists, or reuse of any copyrighted component of this work in other works.
		\end{minipage}
	}
}

\newcommand{\DataItemName}[1]{\textbf{\emph{#1}}}
\newcommand{\DataItemNumber}[1]{\textbf{#1}}

\newcounter{F}
\newcommand{\DataItem}[2]{%
\refstepcounter{F}
\label{#2}
\noindent\DataItemNumber{F\arabic{F}}\ifthenelse{\equal{#1}{\empty}}{}{~\DataItemName{#1}:}}

\newcommand{\refDataItem}[1]{\mbox{F\ref{#1}}}

\title{How\,do\,we\,Evaluate\,Self-adaptive\,Software\,Systems?\\[.35em] \Large A Ten-Year Perspective of SEAMS}

\author{
	\IEEEauthorblockN{
		Ilias Gerostathopoulos$^1$,
		Thomas Vogel$^2$,
		Danny Weyns$^3$, and
		Patricia Lago$^1$
	}
	\IEEEauthorblockA{\textit{$^1$Vrije Universiteit Amsterdam, Netherlands, $^2$Humboldt-Universit\"at zu Berlin, Germany, $^3$KU Leuven, Belgium}}
}

\begin{document}

\textheight = 690pt

\maketitle

\begin{abstract}
With the increase of research in self-adaptive systems, there is a need to better understand the way research contributions are evaluated. Such insights will support researchers to better compare new findings when developing new knowledge for the community. However, so far there is no clear overview of how evaluations are performed in self-adaptive systems. To address this gap, we conduct a mapping study. The study focuses on experimental evaluations published in the last decade at the prime venue of research in software engineering for self-adaptive systems---the International Symposium on Software Engineering for Adaptive and Self-Managing Systems (SEAMS). Results point out that specifics of self-adaptive systems require special attention in the experimental process, including the distinction of the managing system (\ie the target of evaluation) and the managed system, the presence of uncertainties that affect the system behavior and hence need to be taken into account in data analysis, and the potential of managed systems to be reused across experiments, beyond replications. To conclude, we offer a set of suggestions derived from our study that can be used as input to enhance future experiments in self-adaptive systems. 
\end{abstract}

\section{Introduction}\label{sec:Motivation}

Increasingly, we expect software-intensive systems be able to change their structure and behavior at runtime to continue meeting their goals while operating under uncertainty---they need to be self-adaptive. 
Self-adaptation is typically realized via feedback loops that continuously monitor a system and enact changes to the system. 
Self-adaptation has been an active area of research for over 20 years~\cite{weyns2020}, initiated by IBM's pioneering vision of autonomic computing~\cite{kephart2003vision} and the seminal work of Oreizy \etal~\cite{Oreizy1999} and Garlan \etal~\cite{garlan2004rainbow}. 

Numerous new approaches focusing on a variety of aspects of engineering self-adaptive systems (runtime models~\cite{ModelsAtRuntime}, modeling languages~\cite{2555612}, verification at runtime~\cite{8008800}, planning~\cite{0002MCG16}, \etc) have been proposed by the research community. 
To that end, a set of exemplars and reusable artifacts were developed for use by the self-adaptive systems community.\footnote{Exemplars published at SEAMS: \href{http://self-adaptive.org/exemplars}{http://self-adaptive.org/exemplars}.}
Given this substantial body of work in the area, it is important to obtain a clear view of how 
contributions have been evaluated. 
While related work has shed light on some aspects of evaluation, \eg~\cite{1808984.1808985,PEREZPALACIN20141,RAIBULET2017325}, to the best of our knowledge, no study has targeted an in-depth analysis and characterization of the way experimental evaluations have been conducted.

Evaluation is central to self-adaptive systems (as for any type of software systems), since  novel approaches must be assessed based on their contribution~\cite{Assessing-Empirical-Evaluations-2016}. Yet, evaluating contributions of self-adaptive systems may raise particular challenges due to the specifics of these systems~\cite{Roadmap2009} (\eg the use of feedback loops to realize adaptation) and their ability to deal with uncertainty during operation~\cite{Calinescu2020}. Understanding the state of the art in conducting evaluations in self-adaptive systems enables researchers to better compare new findings. Hence, it is important to provide a systematic overview of evaluations of self-adaptive systems, which is currently missing. 

To fill this gap, we performed a mapping study~\cite{PETERSEN20151} aimed to address the question ``How do we evaluate self-adaptive software systems?''  
We focus on experimental evaluations, \ie evaluations that use one or more experiments, since experiments are the most common evaluation approach used in self-adaptive systems. Concretely, the study is centered on (i)~the scope of experiments, (ii)~the way experiments are designed and operated, and (iii)~the way the results of such experiments are analyzed, and (iv) packaged.

The remainder of this paper is structured as follows. 
Section~\ref{sec:focus} presents background and related work.
In Section~\ref{sec:protocol}, we summarize the study protocol, including research questions and process.
Section~\ref{sec:results} presents the results of the study and answers the research questions. 
In Section~\ref{sec:insights}, we discuss insights and threats to validity, and we conclude in Section~\ref{sec:conclusions}.

\section{Background and Related Work}\label{sec:focus}

\subsection{Basic Concepts of Self-Adaptive Systems and Experiments}\label{subsec:SAS}

This study focuses on what is known as architecture-based adaptation~\cite{Oreizy1999,garlan2004rainbow, Kramer2007SMS}, \ie a widely applied approach to realize self-adaptation (see~\cite{Weyns19chap} for an overview). In architecture-based adaptation, a self-adaptive system comprises a \textit{managed system} that is controllable and subject to adaptation, and a \textit{managing system} that performs the adaptations of the managed system. The managed system operates in an \textit{environment} that is non-controllable. 
The managing system forms a feedback loop that is structured according to the MAPE-K reference model, comprising four functions: Monitor-Analyze-Plan-Execute that share Knowledge~\cite{kephart2003vision}. 
In this mapping study, we analyze primary studies from the perspective of architecture-based adaptation and MAPE-K. 

We explain now the basic concepts that we used in the study design. These concepts are based on the process and basic artifacts used in controlled experiments~\cite{Wohlin+2012}. 
While we rely on these concepts, we are interested in all papers that apply an experiment in the broad sense, meaning papers that include most of the stages of the process of controlled experiments, explicitly or implicitly. Our particular focus is on technology-oriented experiments that have systems and software elements as subject of the study (in contrast to studies with humans). 

An experiment starts with an \textit{idea} for an evaluation, for instance, evaluate a new runtime analysis technique and compare it with a state-of-the-art approach. This idea is turned into a hypothesis.\footnote{Instead of hypotheses, researchers may use research questions, or even informal descriptions to capture the idea of the evaluation.} The experiment then tests this hypothesis by studying the effect of manipulating one or more independent variables of the studied case. The three types of independent variables are: constants that have a fixed value for the whole experiment, factors that are expected to have an effect on the outcome, and blocking factors that may have an effect but we are not interested in that effect.\footnote{If an experiment includes a blocking factor whose values create blocking groups, the analysis of the main factors (for which the experiment wants to study the effect) is performed within each blocking group to increase the experiment's precision.}
We use the term \textit{experiment configuration} to refer to the assignment of values to the independent variables\footnote{An experiment configuration relaxes Wohlin \etal's definition of a treatment being the assignment of a particular value to one factor~\cite{Wohlin+2012}.} and \textit{experiment design} to refer to the set of experiment configurations under study. 
During an experiment, the effect on the dependent variables caused by different experiment configurations 
(\ie selected values for the independent variables) 
can be measured. Hence, an experiment essentially tests the relationship between the experiment configurations and the outcome, allowing researchers to draw conclusions about the cause-effect relationship to which the approach under study refers for the stated evaluation problem.

The process of an experiment comprises five steps~\cite{Wohlin+2012}: (1)~\textit{experiment scoping} defines the goals of the experiment; (2)~\textit{experiment planning} refines the goals to determine the experiment design, which includes selecting a context in which the experiment is carried out, formulating the hypothesis to be tested, selecting the independent and dependent variables, selecting subjects, choosing the experiment configurations,
defining how the experiment should be executed and monitored, and evaluating the validity of the results; (3)~\textit{experiment operation} prepares and executes the experiment, (4)~\textit{analysis \& interpretation} analyzes the data collected from the experiment and tests the hypothesis, and (5)~\textit{presentation \& packaging} presents the results and makes a replication package available. 

\subsection{Focus of our Study}
\label{subsec:Review-Focus}

This mapping study aims at understanding how evaluations of self-adaptive software systems are performed in studies presented at the International Symposium on Software Engineering for Adaptive and Self-Managing Systems (SEAMS). To focus the review, we performed a preliminary analysis of the evaluation methods that were applied in the full papers published at SEAMS between 2011 and 2020. 
We labelled the evaluation methods according to the following categories: no evaluation, showcase, experiment, review, questionnaire, and proof.
A showcase presents results from a \textit{single} experiment configuration~\cite[p.\,75]{Wohlin+2012}.
An experiment, on the other hand, provides \textit{quantitative} comparative results \textit{for more than one} experiment configuration.
A controlled experiment is an experiment that follows a rigorous well-defined process~\cite{Wohlin+2012}. 
We found that more than 65\% of the examined studies (82 out of 126 full papers) contained at least one experiment.
Since the majority of studies use experiments for evaluation, we decided to focus our study on experiments as the evaluation method. 

\subsection{Related Work}\label{subsec:related_work}

In the field of self-adaptation, several related efforts pay attention to contributions in the field~\cite{Krupitzer2015,Patikirikorala2012,Muccini2016,MAHDAVIHEZAVEHI20171,DAngelo2019,Grua2019}, but do not provide an in-depth study of evaluation aspects. Other related studies do consider evaluation aspects, but they take a specific angle focusing on: claims and evidence in self-adaptive systems~\cite{weyns2013claims,weyns2012claims}, quality aspects and metrics~\cite{978-3-540-30207-0_74,1808984.1808985,1988008.1988020,PEREZPALACIN20141,RAIBULET2017325,deSousa+2019}, and methodology~\cite{Andersson2013,Porter2020,Taranu+2010}. In contrast, our mapping study targets an in-depth analysis and characterization of the way experimental evaluations have been designed, conducted, analyzed, and packaged.

\section{Summary of the Protocol}\label{sec:protocol}

Following the guidelines of~\cite{PETERSEN20151}, we conducted the mapping study with four researchers that jointly developed the protocol. To ensure  validity, the protocol was also reviewed by experts in self-adaptation and experimental software engineering. 
We made the protocol available as part of our replication package.\footnote{\url{https://doi.org/10.5281/zenodo.4622749}}

\subsection{Research Questions}\label{sec:rqs}

We formulate the goal of our mapping study by using the classic Goal-Question-Metric (GQM) approach~\cite{van2002goal}:
\begin{quote}
\textit{Purpose}: Organize and characterize \\
\textit{Issue}: how evaluations of self-adaptive software systems are performed \\
\textit{Object}: in research on self-adaptation published at the 10 most recent SEAMS installments \\
\textit{Viewpoint}: from a researcher’s viewpoint. 
\end{quote}

We detailed this goal into five concrete research questions that correspond to the five phases of the experiment process.

\begin{description}

\item[RQ1:] What is the scope of experiments? 

\item[RQ2:] What is the experimental design? 

\item[RQ3:] How are experiments operated?

\item[RQ4:] How is the experiment data analyzed?

\item[RQ5:] How are experiment results packaged?

\end{description} 

With RQ1, we aim to understand the purpose and object of evaluations.
With RQ2, we want to obtain in-depth insights in independent and dependent variables, experiment configurations, and designs. 
This will shed light on the complexity and variability of experiments applied for self-adaptive systems. 
With RQ3, we want to characterize how experiments on self-adaptive system are executed, with particular emphasis on  aspects specific to self-adaptive systems such as the distinction between managed and managing system.
With RQ4, we want to get insights of how experiment results are analyzed (\eg using descriptive or inferential statistics).
Finally, with RQ5, we want to obtain an overview of whether and how experiment results are made available and packaged for replication.

\subsection{Search Strategy}\label{sec:search}

We examine primary studies published at the main venue on engineering self-adaptive systems---SEAMS.
First, there is a normative justification for this focus. Studies presented at SEAMS provide a representative sample of software engineering research of self-adaptive systems. Other studies have also chosen to focus on one specific venue such as ICSE~\cite{Sousa2019,Zannier2006}. 
According to the ACM SIGSOFT Empirical Standards~\cite{ralph2020acm}, which are currently under development, this is an acceptable deviation to perform secondary studies. 
Second, there is a qualitative justification. To make a useful and accurate assessment of the features we target in this review, we need relevant data. Based on our combined experience as active members of the SEAMS community, we believe that studies presented at SEAMS provide a source of such relevant data. In light of these two arguments, we acknowledge that our focus may create some degree of bias that we further discuss as a threat to external validity.

\subsection{Inclusion and Exclusion Criteria}\label{sec:inclusion-exclusion}

We use the following inclusion criteria to select papers:

\begin{itemize}
    \item IC1: The paper is published at SEAMS between 2011 and 2020 (included). In 2011 SEAMS became a symposium, which increased the level of the evaluations significantly.

    \item IC2: The paper empirically evaluates an approach by using one or more technology-oriented experiments. 
\end{itemize}

\noindent We use the following exclusion criteria:

\begin{itemize}
\item EC1: The paper is not a full research paper. These papers typically do not empirically evaluate a new approach.

\item EC2: The paper presents a secondary study (\eg~literature review, survey, mapping study) or an overview of the field (\eg~taxonomy, roadmap). These papers do not present and evaluate a novel approach for self-adaptation.
\end{itemize}

A paper is selected if it meets all of the inclusion criteria and does not meet any exclusion criterion.

\subsection{Data Items}\label{sec:extraction}

To answer the research questions, we define a set of data items to be extracted from the papers, see  Table~\ref{tab:data_items}.
Since the data items refer to a single experiment and a study may contain more than one experiment, we identify all the experiments that are included in a study and then extract the data of each experiment independently. 
The column ``Process Step'' shows that our study covers the whole experiment process (see Section~\ref{subsec:SAS}) and key aspects that are relevant for technology-oriented experiments as
reported at SEAMS.

\renewcommand{\arraystretch}{1.1}

\begin{table*}[!ht]
\centering
\caption{Extracted data items.}
\label{tab:data_items}
\begin{tabular}{l p{4cm} l p{2.9cm} p{8.2cm}} \toprule
\hspace{3pt}\textbf{ID} & \textbf{Item} & \textbf{Use} & \textbf{Process Step} & \textbf{Explanation}\\ \midrule
 \DataItem{}{EvaluationTarget}  & Target of evaluation       & RQ1 & Experiment scoping &  The main element that is subject of evaluation, incl. the whole feedback loop and methods for distinct MAPE-K stages and learning. 
 \\
 \DataItem{}{Objectives}      & Objectives of evaluation       & RQ1      & Experiment scoping &  The aspects of the proposed approach that are evaluated, mentioned explicitly or implicitly. 
 \\ \midrule
 \DataItem{}{EvaluationQuestion}      & Formulation of evaluation problem       & RQ2      & Experiment planning &  Captures whether there is an explicit formulation of the evaluation problem by either research questions or hypotheses.   
 \\ 
 \DataItem{}{ConstantIndVariables}      & Constants  & RQ2   & Experiment planning &  The names of the variables that are constant across the experiment. 
 \\ 
 \DataItem{}{BlockingFactors}  & Blocking factors       & RQ2    & Experiment planning & The names of the variables that are used to create experiment blocks, but without interesting effect~\cite[p.\,94]{Wohlin+2012}. 
 \\ 
  \DataItem{}{Factors}      & Factors       & RQ2    & Experiment planning &  The names of the variables that change across experiment configurations. 
 \\ 
 \DataItem{}{DepVariables}  & Dependent variables  & RQ2     & Experiment planning &  The names of the variables that measure the effect of an experiment configuration, also called ``response variables''~\cite[p.\,74]{Wohlin+2012}).    
 \\ 
  \DataItem{}{ExpTreatments} & Counts experiment variables & RQ2    & Experiment planning &   The number of values of independent and dependent variables used in experiments (referring to \refDataItem{ConstantIndVariables},  \refDataItem{BlockingFactors}, \refDataItem{Factors}, and \refDataItem{DepVariables}). 
  \\ 
  \DataItem{}{DesignType}    & Design type   & RQ2  & Experiment planning &   The design type used in the experiment, following the standard design types described by Wohlin \etal~\cite[p.\,95]{Wohlin+2012}.   
  \\ \midrule
 \DataItem{}{ManagedSystem} & Managed system name & RQ3  &  Experiment operation  & The name of the managed system, if any. The managed system may be a SEAMS artifact, see  \href{http://self-adaptive.org/exemplars}{http://self-adaptive.org/exemplars}. 
 \\ 
 \DataItem{}{ManagedSystemNature}      & Nature of managed system       & RQ3   &  Experiment operation &  The type of managed system used in the evaluation, incl. model, simulated/emulated, real implementation, and real-world application. 
 \\ 
 \DataItem{}{DataProvenance}      & Data provenance       & RQ3   &  Experiment operation &  Source of data related to the users or the environment of the managed system, incl. synthetic data, emulated data, and real-world data.
 \\ 
 \DataItem{}{Uncertainty}      & Uncertainty       & RQ3    &  Experiment operation &  
 	The way uncertainty is represented in the experiment. This type of uncertainty can create the need for self-adaptation. 
 \\ \midrule
 \DataItem{}{AnalysisType}      & Type of analysis       & RQ4    &  Analysis \& interpretation &  The type of analysis that is performed on the experiment results, incl. none, exposition (narrative), descriptive statistics, and statistical tests. 
 \\ 
 \DataItem{}{AnswerToEvaluationProblem}      & Answer to evaluation problem       & RQ4    & Analysis \& interpretation &  	Whether there is an explicit answer to research questions or hypotheses. 
 \\ 
 \DataItem{}{ValidityThreats}      & Threats to Validity    & RQ4    & Analysis \& interpretation &    The types of threats to validity mentioned (in a dedicated section/subsection or paragraph), if any. 
 \\ \midrule
 \DataItem{}{ResultsAvailable}      & Results available       & RQ5    & Presentation \& packaging &   Captures whether evaluation results are available (\eg via a URL). 
 \\ 
 \DataItem{}{CodeAvailable}      & Degree of reproducibility       & RQ5  & Presentation \& packaging & Captures whether the implementation of the approach is available or the full replication package is available (\eg via a URL).
 \\
 \bottomrule
\end{tabular}
\end{table*}

\subsection{Approach for the Analysis}

We tabulate the data that we extract from the primary studies in spreadsheets for processing. We use descriptive statistics to structure and present the quantitative aspects of the data, and comprehensible summaries of the data to answer the research questions. We present results with plots using simple numbers and sometimes means and standard deviations to help understand the results. 
For the data items 
\refDataItem{EvaluationTarget}, \refDataItem{Objectives},
\refDataItem{ConstantIndVariables}--\refDataItem{DepVariables}, and 
\refDataItem{Uncertainty},
we collect free text and apply coding~\cite{Vollstedt2019} to capture the essence of the answers. 
As a concluding step, we produce a schematic overview of the experimental process for self-adaptive systems, allowing us to identify any difference from the ``traditional'' experiment process (see Section~\ref{subsec:SAS}).

\section{Results}\label{sec:results}

\subsection{Demographic Information}

\begin{figure}[h!]
\centering
\includegraphics[width=1\columnwidth]{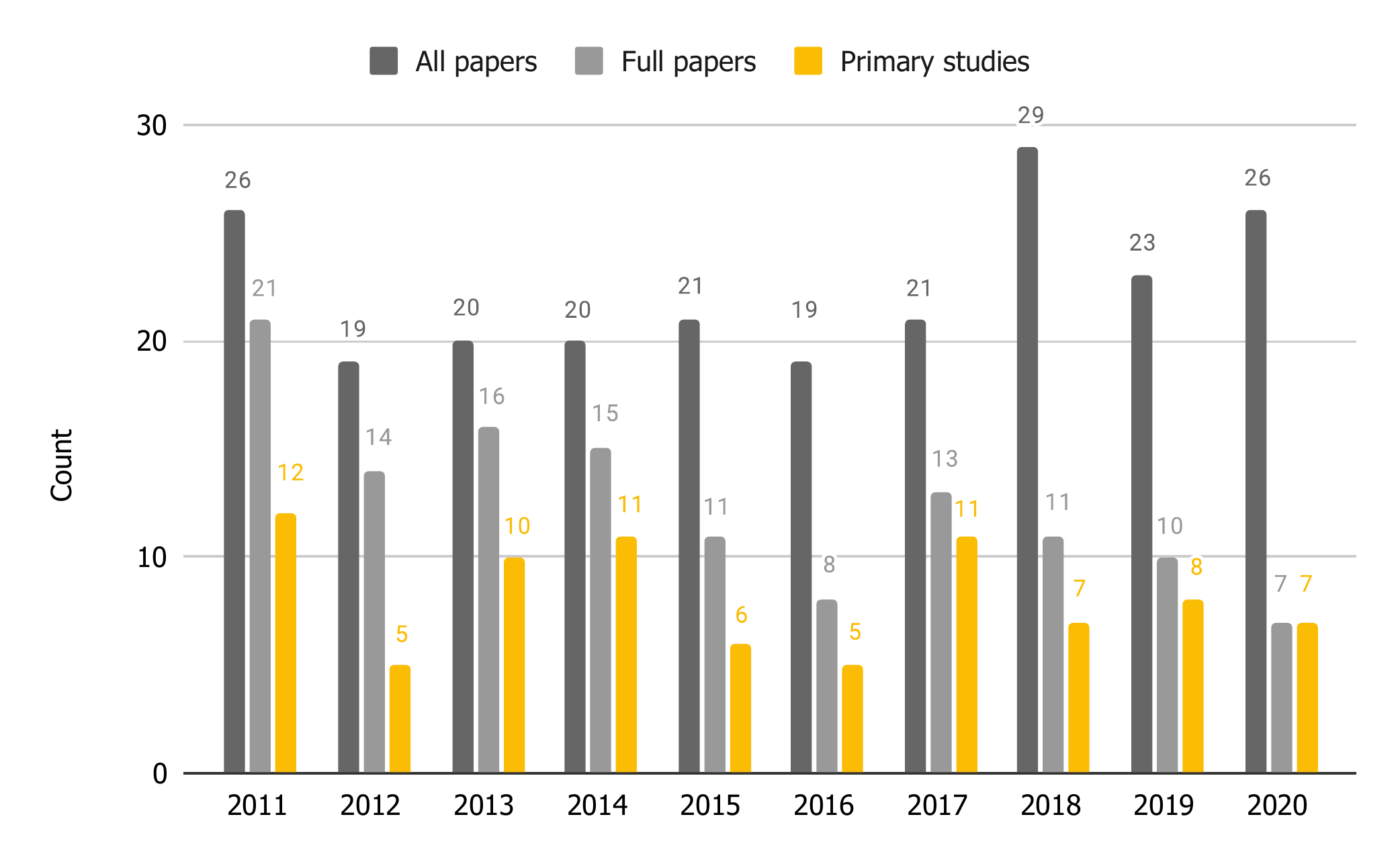}
\caption{Overview of selected primary studies per year.}
\label{fig:primary_studies_per_year}
\vspace{-.5em}
\end{figure}

From a total of 224 papers presented at SEAMS in the period between 2011 and 2020, we identified 126 full papers and from those, we selected 82 primary studies (65\%) after applying the inclusion and exclusion criteria, see Fig.~\ref{fig:primary_studies_per_year}. The primary studies reported a total of 140 experiments (1.71 on average, 0.34 std.). The results show that the relative number of full papers that use experiments for evaluation increased from 57\% in the period from 2011 to 2015 to 78\% in the period from 2016 to 2020 (with even 100\% in 2020). At the same time, the average number of experiments per primary study increased from 1.52 in the first period to 1.92 in the second period. These numbers underpin an increasing level of mature evaluations in papers published at SEAMS over time.

\subsection{What is the Scope of Experiments?}

To answer RQ1, we collected the data about the targets of the evaluations  (\refDataItem{EvaluationTarget}), and the objectives of evaluations (\refDataItem{Objectives}). 

\begin{figure}[t]
\centering
\includegraphics[width=1\columnwidth]{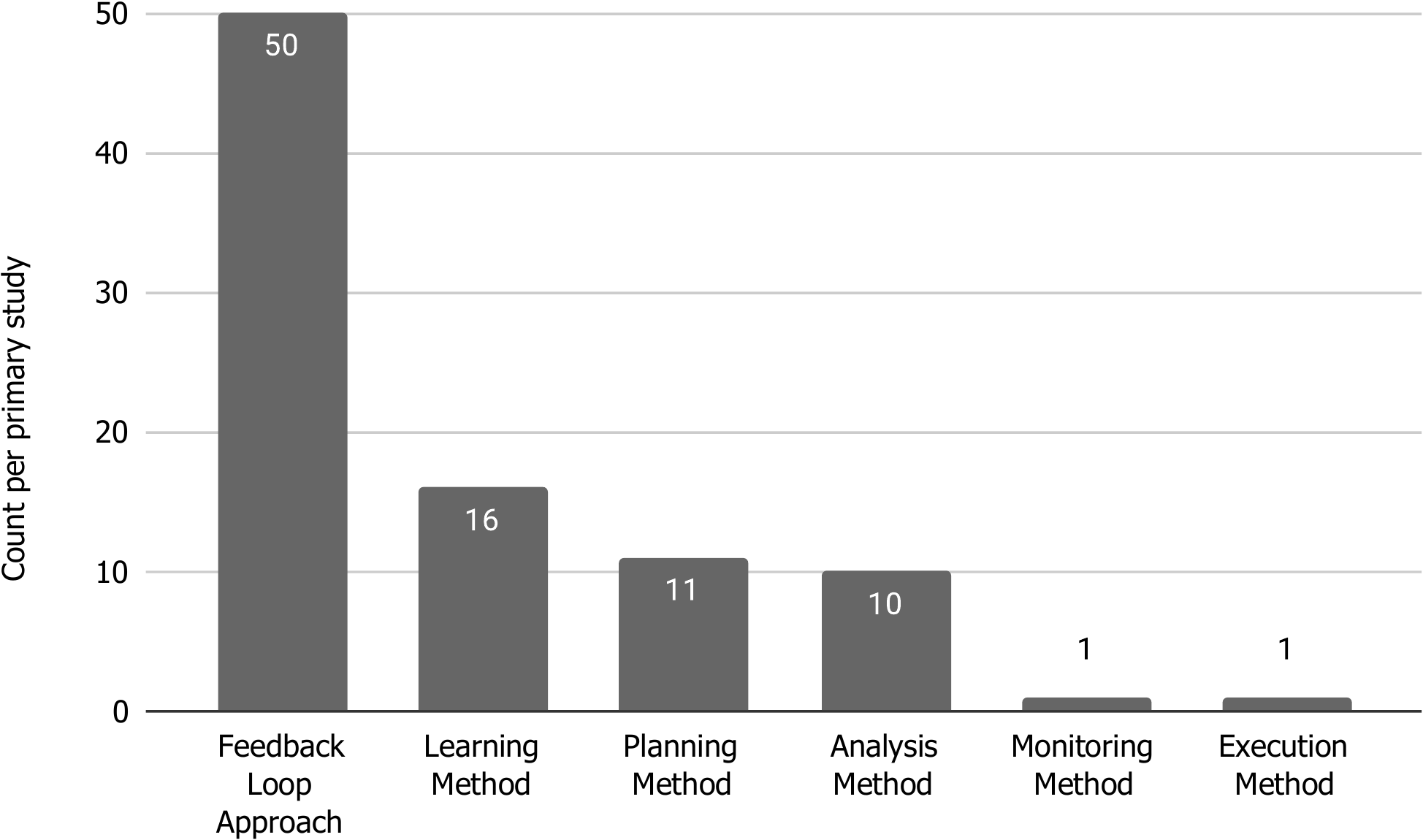}
\caption{Count of evaluation targets (\refDataItem{EvaluationTarget}) per primary study.}
\label{fig:contributions}
\vspace{-10pt}
\end{figure}

Fig.~\ref{fig:contributions} plots the counts for the targets of the evaluations~(\refDataItem{EvaluationTarget}) reported in the primary studies. In 50 studies (61\%), the evaluation targeted a new integrated adaptation approach that covers the full feedback look. For instance, Derakhshanmanesh \etal~\cite{Derakhshanmanesh+2011} present an adaptation framework that uses graph-based models throughout the feedback loop. Next, 16 studies (20\%) evaluated a new learning method. For instance, Duarte \etal~\cite{Duarte2018} contribute a method for learning linear models that capture non-deterministic impacts of adaptation. 
Notably, the numbers for new learning methods increased from five in the period from 2011 to 2015 to 11 in the period from 2016 to 2020. 
The remaining studies focused on evaluating new approaches for distinct MAPE-K stages. Among these, 11 studies targeted a planning method and ten studies targeted an analysis method. Only one study targeted a new monitoring approach~\cite{Mertz+2019} and one other study targeted an execution approach~\cite{Gambi+2013}. 
Independently of the evaluation targets, 79 of all 140 experiments (56\%) reported in the primary studies used the full feedback loop, while 61 experiments (44\%) considered only a part of the feedback loop in the evaluation.

\begin{figure}[t]
\centering
\includegraphics[width=1\columnwidth]{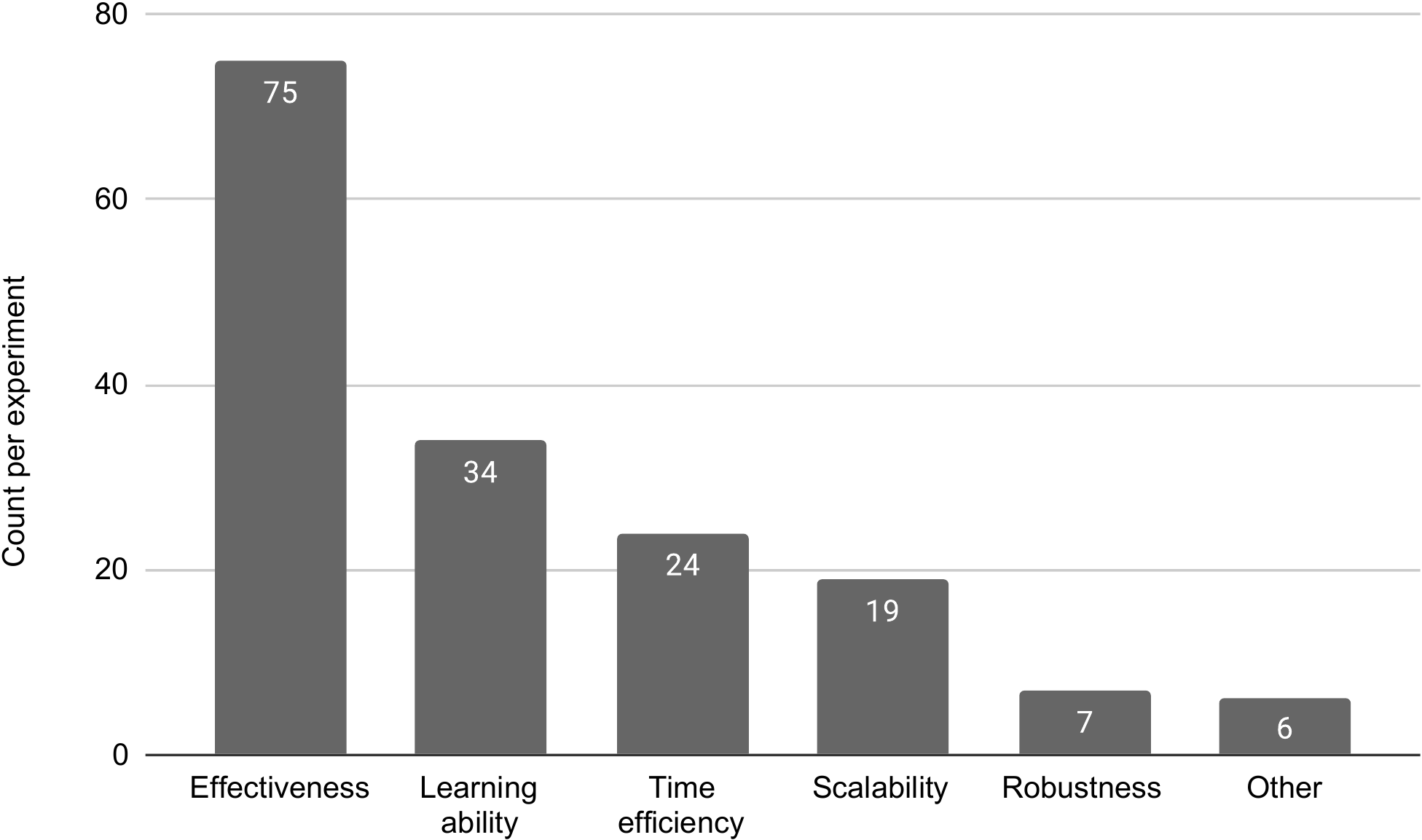}
\caption{Count of evaluation objectives (\refDataItem{Objectives}) per experiment.}
\label{fig:evaluation_objectives}
\end{figure}

Fig.~\ref{fig:evaluation_objectives} shows the results of the evaluation objectives~(\refDataItem{Objectives}) used in the experiments. We extracted 165 individual evaluation objectives from the 140 experiments: 116 experiments (83\%) had one evaluation objective, 23 (16\%) had two objectives, and one had three objectives. The top-reported evaluation objective is effectiveness that was used 75 times (45\% of 165), followed by learning ability (used 34 times, 21\%) and time efficiency (used 24 times, 15\%).  
As examples, Jamshidi~\etal evaluate the effectiveness in terms of the number of completed robot missions~\cite{Jamshidi2019}, while Nikravesh \etal evaluate the learning ability by assessing accuracy of different workload predictors~\cite{Nikravesh+2015}. 
Sousa \etal evaluate the time efficiency of planning by measuring the time to find a valid configuration~\cite{Sousa+2017}, while Shin \etal evaluate the scalability of a search-based adaptation approach in terms of execution time with increasing network size~\cite{Shin+2020}.

\begin{figure}[t]
\centering
\includegraphics[width=1\columnwidth]{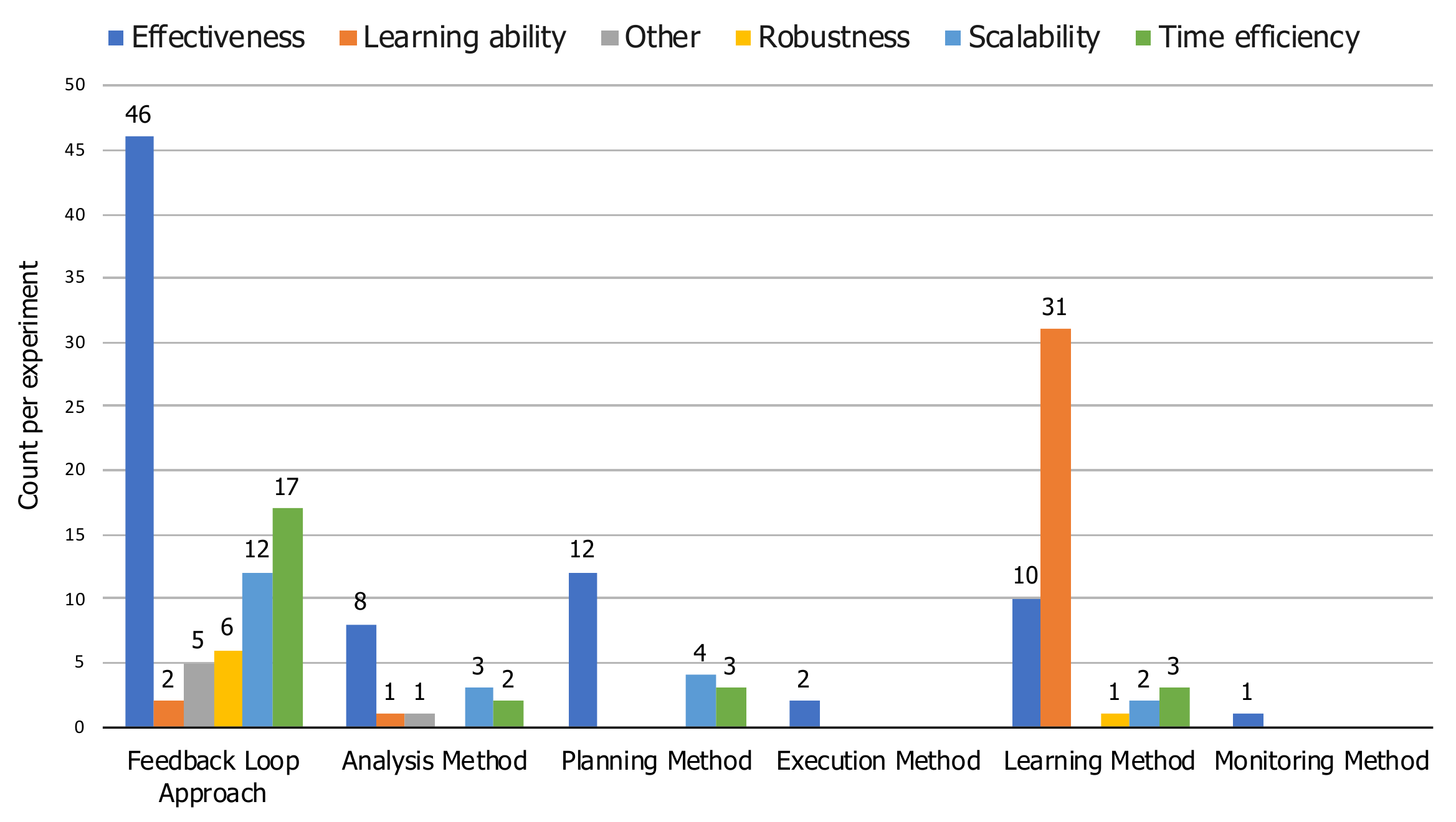}
\caption{Mapping of evaluation targets (\refDataItem{EvaluationTarget}) to objectives (\refDataItem{Objectives}).}
\label{fig:evaluation_objectives_to_targets}
\end{figure}

Fig.~\ref{fig:evaluation_objectives_to_targets} maps the evaluation targets~(\refDataItem{EvaluationTarget}) to the  objectives~(\refDataItem{Objectives}). The results show that effectiveness is used as evaluation objective for all types of evaluation targets. New feedback loop approaches are mostly evaluated for effectiveness (46 experiments) and time efficiency (17 experiments). 
Not surprisingly, learning ability is the top evaluation objective for new learning methods (in 31 experiments). 
Scalability is used as evaluation criterion for four of the six evaluation targets (not for the single new proposed execution~\cite{Gambi+2013} and monitoring methods~\cite{Mertz+2019}).

\vspace{5pt}
\begin{tcolorbox}[size=small]
\small 
\textbf{Answer to RQ1: What is the Scope of Experiments?} 
The main evaluation target of experiments in self-adaptive systems is a new integrated feedback loop approach with effectiveness and time efficiency as main evaluation objectives. Recently, we observe a rapid increase in experiments that focus on new learning approaches evaluated for their ability to learn and effectiveness.  
\end{tcolorbox}

\subsection{What is the Experimental Design?}

To answer RQ2, we collected data about the formulation of the evaluation problem (\refDataItem{EvaluationQuestion}), the independent variables (\ie constants  (\refDataItem{ConstantIndVariables}), blocking factors (\refDataItem{BlockingFactors}), and factors (\refDataItem{Factors})),  the dependent variables (\refDataItem{DepVariables}), the counts of values of the different types of variables (\refDataItem{ExpTreatments}), and the design type (\refDataItem{DesignType}), see Table~\ref{tab:data_items}.  

Only 28 studies (34\%) provide a well-defined formulation of the evaluation problem (\refDataItem{EvaluationQuestion}), of which 21 (26\%) use research questions and 7 (8\%) use hypotheses. 
For example, Jamshidi \etal specify three research questions (on accuracy, effectiveness, and robustness) that guide their evaluation~\cite{Jamshidi+2014}, while Fredericks uses null hypotheses to compare the proposed approach to a baseline (on effectiveness of generating adversary environments and effectiveness of adaptation)~\cite{Fredericks2016}. 
The remaining 54 primary studies (66\%) provide an informal description of the evaluation problem.\footnote{With informal description we mean the evaluation problem is described with some general words or is only provided implicitly.} 
For example, Pournaras \etal describe the goal of their evaluation in an informal way~\cite{Pournaras+2018}. 
Remarkably, this result resembles those of an early survey of primary studies of SEAMS before the year 2012~\cite{weyns2012claims}, suggesting little progress in formulating clearly defined research problems in studies presented at SEAMS. 

\begin{figure}[t]
\centering
\includegraphics[width=1\columnwidth]{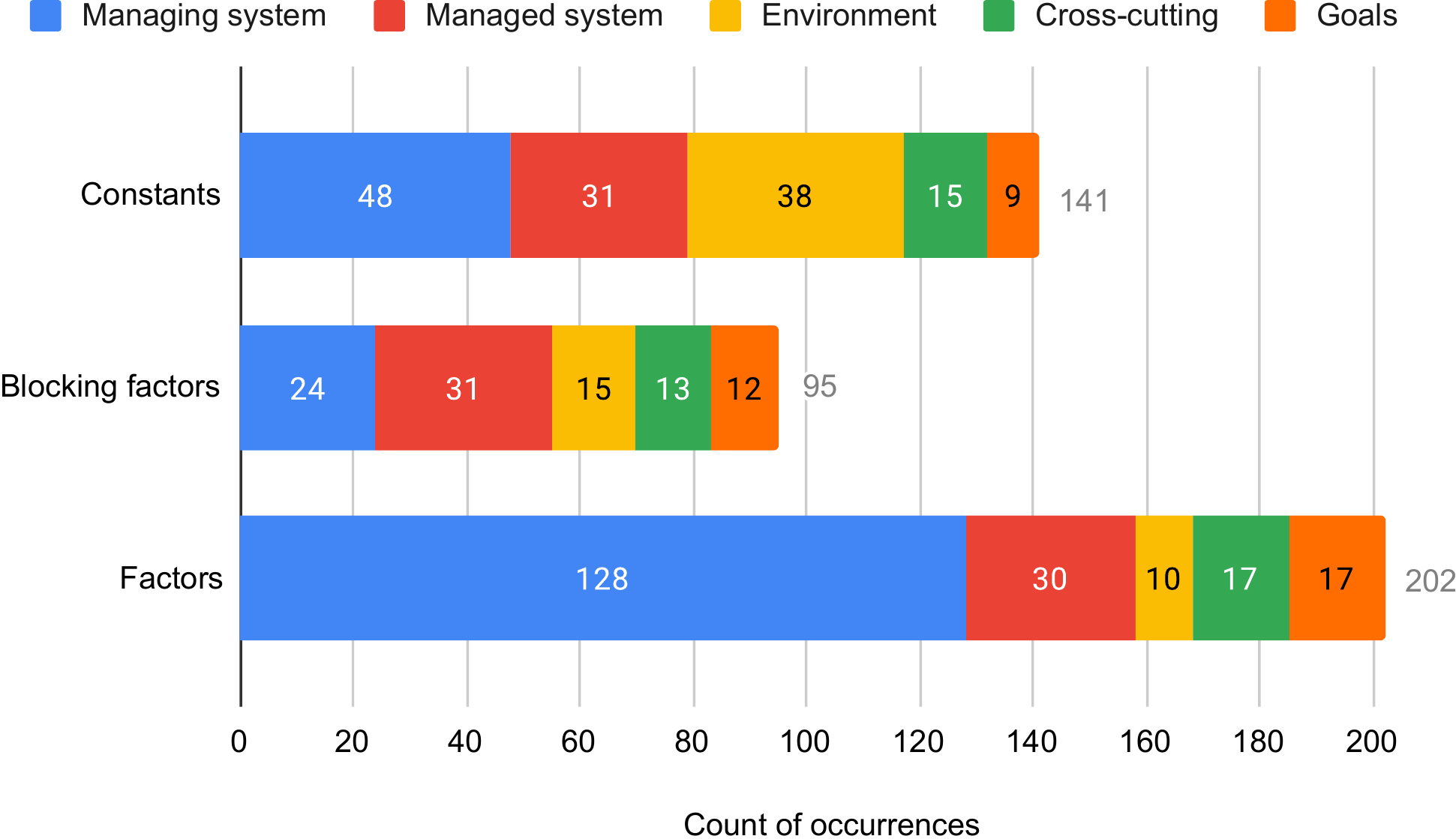}
\caption{Count of independent variables (\refDataItem{ConstantIndVariables}, \refDataItem{BlockingFactors}, \refDataItem{Factors}).}
\label{fig:independent_variables}
\vspace{-.5em}
\end{figure}

Fig.~\ref{fig:independent_variables} gives an overview of the overall count of independent variables (\refDataItem{ConstantIndVariables}, \refDataItem{BlockingFactors}, \refDataItem{Factors}) in the experiments (see Section~\ref{subsec:SAS} for a description of the different types of independent variables). 
From all the experiments reported in the primary studies we extracted 141 constants (avg. per experiment 1.01, std. 0.95). 
``Load profile" is the constant with the highest number of occurrences (used in 26 experiments, \eg ~\cite{DiMarco+2013,Huang+2013,experiment-example}); other examples are ``number of nodes"~\cite{Chen+2014}, ``number of sensors"~\cite{Gerasimou+2014} and ``learning/optimization hyperparameters"~\cite{Pascual+2013,Nikravesh+2015}.
From all experiments, we extracted 95 blocking factors (avg. per experiment 0.68, std. 0.91). For example, ``deployment environment" was used to block the analysis of elasticity configurations in two settings: private and public cloud~\cite{Herbst+2015}.
Finally, we extracted in total 202 factors (avg. per experiment 1.44, std. 0.85). For example, ``assurance approach" in~\cite{Fredericks2016} took two values that correspond to the proposed approach (genetic algorithm) and a baseline (random search) that are evaluated against adversarial environments of the system.

The results show that 200 of a total of 438 independent variables (46\%) relate to the managing system. Specifically, 128 of a total of 202 factors (28\%) relate to the managing system, indicating that experiments in self-adaptive systems target prominently the evaluation of new approaches and methods of the managing system. 
On the other hand, 92 independent variables (21\%) relate to the managed system and 63 relate to the environment (14\%), the latter are mostly constants in the experiments. 
Notably, only 38 independent variables (9\%) relate to system goals (17 of these are factors). These figures suggest a relatively low interest in considering the impact of goals in the evaluation of new approaches for self-adaptive systems. 
Finally, the group ``Cross-cutting'' refers to independent variables that cross-cut at least two elements of a self-adaptive system (managing system, managed system, environment, goals). We extracted 45 such independent variables (10\%). Among these, the most frequent combination is a variable that cross-cuts managed system and goals representing a scenario that warrants adaptation. 
For example, Shevtsov \etal~\cite{Shevtsov+2017} uses a scenario that is defined by a set of sensors that need to perform a monitoring task with maximum measurement accuracy while being exposed to failures.  

\begin{figure}[t]
\centering
\includegraphics[width=1\columnwidth]{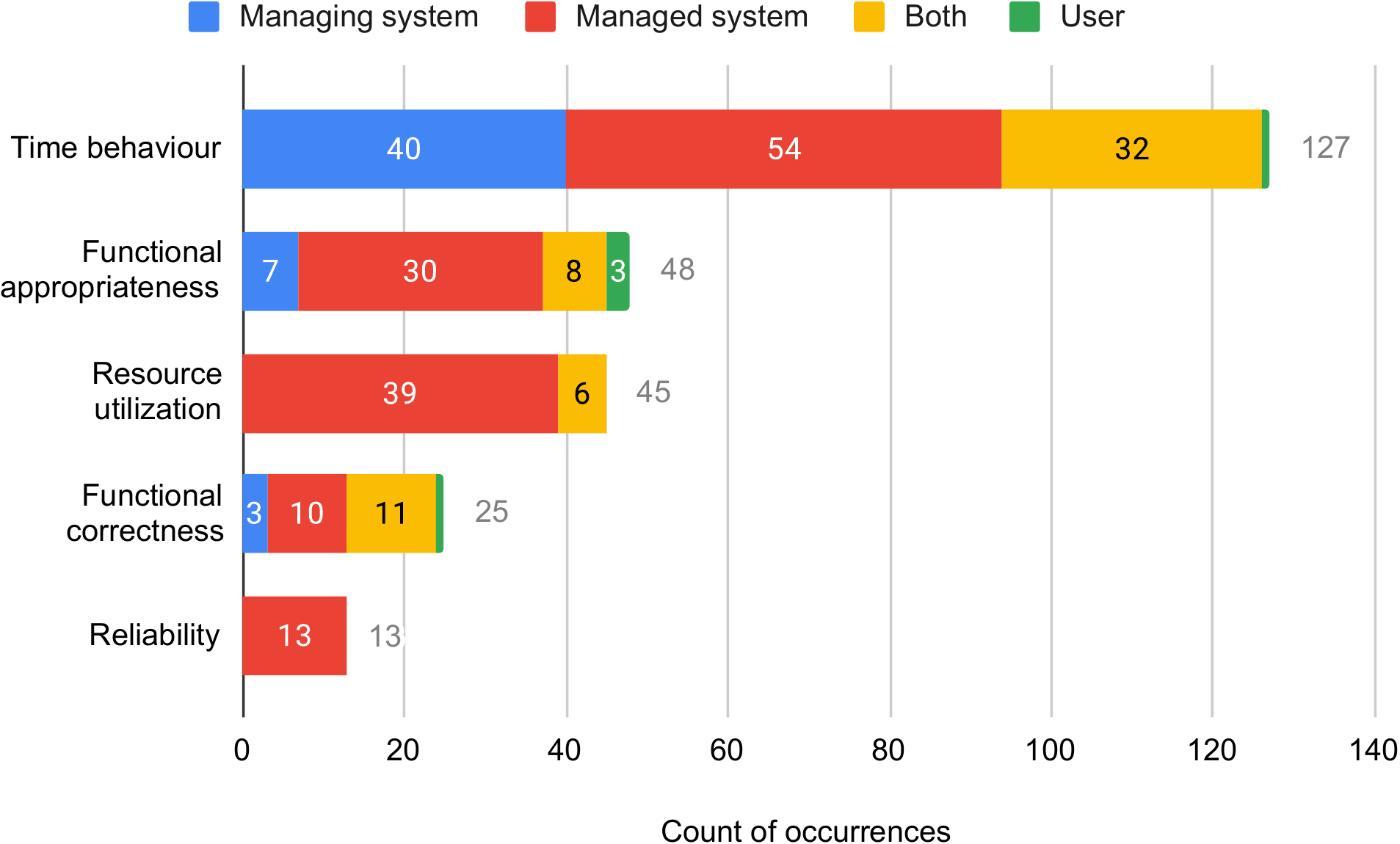}
\caption{Count of dependent variables (\refDataItem{DepVariables}). Out of the total of 267 occurrences, 9 belong to smaller categories not depicted.}
\label{fig:dependent_variables}
\end{figure}

Fig.~\ref{fig:dependent_variables} shows the results for the dependent variables~(\refDataItem{DepVariables}). 
We identified five classes of dependent variables from a total count of 267 concrete variables used in the experiments (on average, 1.91 variables per experiment; 1.10 std.). The dominant dependent variable is ``Time behavior''\footnote{These variables measure a time-related property. The most prominent variables are response time (31 times used) and processing time (29 times).} that was used 127 times (48\% of the total count). As an  example, the response time of an online news service (ZNN, a popular SEAMS artifact) was measured in~\cite{Angelopoulos+2013}.  
Other frequently used dependent variables are ``Functional appropriateness''\footnote{These dependent variables measure the suitability of a new approach from a functional viewpoint. Examples are the degree that goals are satisfied and the degree of financial profit obtained from applying a new approach.} (48 times; 18\% of the total count) and ``Resource utilization'' (45 times; 17\% of the total count). 
For example, ``distance scanned" was used to assess the functional appropriateness of the proposed solution in~\cite{Shevtsov+2017}, while the number of servers was used to assess the resource consumption in~\cite{Herbst+2015}. 
Notably, we found only 13 concrete variables related to ``Reliability.''\footnote{We counted the variables that explicitly refer to reliability, or are clearly connected such as packet loss in a network. However, variables of other classes may indirectly relate to reliability. E.g., a variable that measures functional correctness may be important to achieve a required level of reliability.}
For example, packed loss is used to assess reliability in~\cite{Donckt+2020}.
The dependent variables refer mostly to the managed system (146 times, 55\% of the total count) followed by both managing and managed system (57 times, 21\%) and managing system (50 times, 19\%).

Table~\ref{tab:counts_values_variables} summarizes the average numbers of different types of variables (\refDataItem{ExpTreatments}) used per experiment.

\begin{table}[!ht]
\vspace{-.5em}
\centering
\caption{Average numbers of variables per experiment.}
\label{tab:counts_values_variables}
\begin{tabular}{lll} \toprule
\textbf{Variable} & \textbf{Avg} & \textbf{Std}\\ \midrule
Constants & 1.01 & 0.95  \\ 
Blocking factors & 0.68 & 0.91 \\  
Factors & 1.44 & 0.85 \\  
Dependent variables & 1.91 & 1.10 \\  
\bottomrule
\end{tabular}
\end{table}

\begin{figure}[t]
\centering
\includegraphics[width=1\columnwidth]{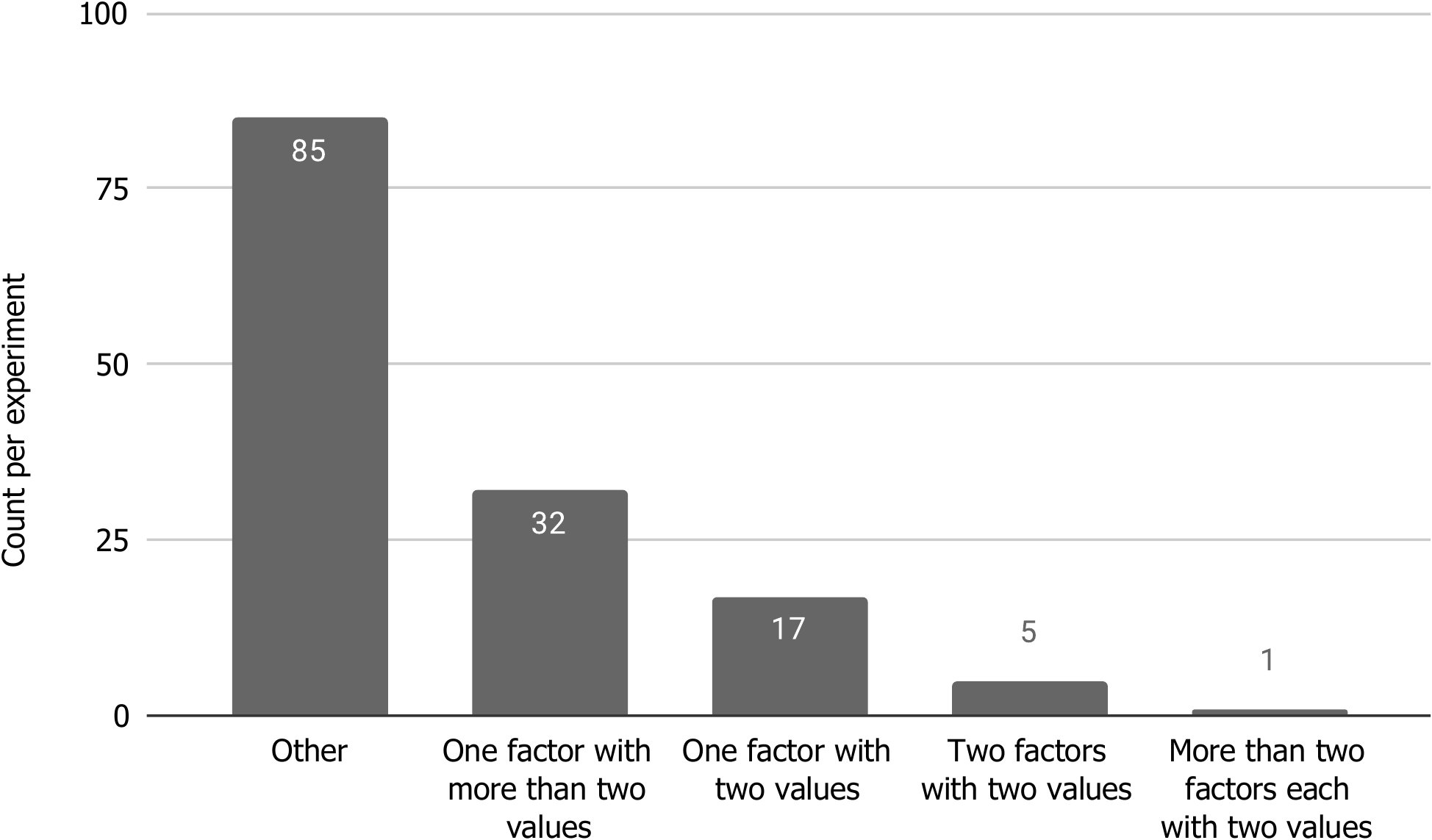}
\caption{Standard design types used in experiments (\refDataItem{DesignType}).}
\label{fig:design_type}
\end{figure}

While these numbers give us an insight in the variables used in the experimental design, we also extracted data about the use of standard design types (\refDataItem{DesignType}) to get a better view of the concrete design types used in experiments in self-adaptive systems.\footnote{We use the four standard design types for experimentation in software engineering described by Wohlin \etal~\cite[p.\,95]{Wohlin+2012}.} 
Fig.~\ref{fig:design_type} shows the results. 
Out of all 140 experiments, 55 experiments (39\%) use a standard design type. 
The most frequently used standard design type is ``One factor with more than two values'' (32 experiments, 23\%). 
For example, the experiment design in~\cite{Camara+2013} involves one factor (``managing system") with three values corresponding to using built-in adaptation mechanisms, using architecture-based adaptation (Rainbow) with default adaptation strategies, and using Rainbow with improved adaptation strategies. 
Of the 55 experiments that use a standard design, 49 (35\% of all experiments) use one factor with two or more values. 
However, a majority of 85 experiments (61\%) do not use a standard design type. 
These experiments use a design with different combinations of factors and values for these factors. 
For instance, the experiment by Kistowski \etal to evaluate load extraction methods has one factor and three blocking factors, generating a total of 96 experiment configurations~\cite{Kistowski+2015}. 

\begin{table*}[t]
\centering
\caption{Identified patterns for different objectives that map independent to dependent variables.}
\label{tab:patterns}
\begin{tabular}{lp{7cm}ll} \toprule
\textbf{Objective} & \textbf{Independent variables} & \textbf{\#} & \textbf{Dependent variables (top two)}\\ \midrule
Effectiveness & Managing-Method (2) $\times$ * (1) & 13 & Resource utilization [managed], Time behaviour [managed]\\ 
Scalability & Managing-Method (1) $\times$ 
Managed-Variation ($>$2) $\times$ * (1) & 7 & Time behaviour [managing], Functional appropriateness [both]\\  
Time efficiency & Managing-Method ($>$2) $\times$ * (1) &  4 &  Time behaviour [both], Time behaviour [managing] \\  
Learning ability & 
Managing-Method (2) $\times$ Managing-Parameter ($>$2) $\times$ * (1) & 3 & Time behaviour [managed], Resource utilization [managed] \\  
\bottomrule
\end{tabular}
\end{table*}

To get further insight in the concrete designs of experiments in self-adaptive systems, we combined the data collected for the evaluation objectives (Fig.~\ref{fig:evaluation_objectives}), independent variables (Fig.~\ref{fig:independent_variables}), and dependent variables (Fig.~\ref{fig:dependent_variables}).
This allowed us to identify a number of patterns for different evaluation objectives that map independent to dependent variables, see Table~\ref{tab:patterns}.\footnote{As notation to describe a combination of independent variables we use: ``$var_1$ ($n_{var_1}$) $\times$ $var_2$ ($n_{var_2}$) $\times$ ...'', where $var_1$ is the variable and $n_{var_i}$ is the number of values of $var_i$. We use an asterisk as a wild card for $var_i$. Note that $n_{var_i}=1$ if $var_i$ is a constant.} 

The pattern for effectiveness applies two methods of a managing system combined with constants, and measures the effect on resource utilization or time behavior of a managed system (13 instances). 
For example, Barna \etal~\cite{Barna+2012} evaluate the effectiveness of mitigating DoS attacks by comparing two mitigation methods based on measured CPU utilization and response time of the managed system. 
The pattern for scalability applies a single method of a managing system on different variations of a managed system, and measures the scalability for time behavior of the managing system or functional appropriateness of both the managed and managing system (8 instances). 
For example, Incerto \etal~\cite{Incerto+2016} evaluate the scalability of an SMT-backed planning approach in terms of computation time under increasing numbers of servers in the managed system. 
The pattern for time efficiency applies more than two methods of a managing system combined with constants, and measures the effect on the time behavior of both managing and managed system or just the managing system (4 instances). 
For example, Kumar \etal~\cite{Kumar+2020} compare the time efficiency of four self-adapting service composition approaches by measuring the planning time. 
Finally, the pattern for learning ability applies two methods of the managing system combined with multiple parameter settings of the managing system (typically related to learning hyperparameters) and constants. It measures the effect on the learning ability (\eg in terms of accuracy or correlation) for the time behavior or functional correctness of a managed system (3 instances). 
For example, Duarte \etal~\cite{Duarte2018} evaluate the accuracy of two learning methods that predict response time of the managed system, configured with different sizes of training data.

\begin{figure}[t]
\centering
\includegraphics[width=1\columnwidth]{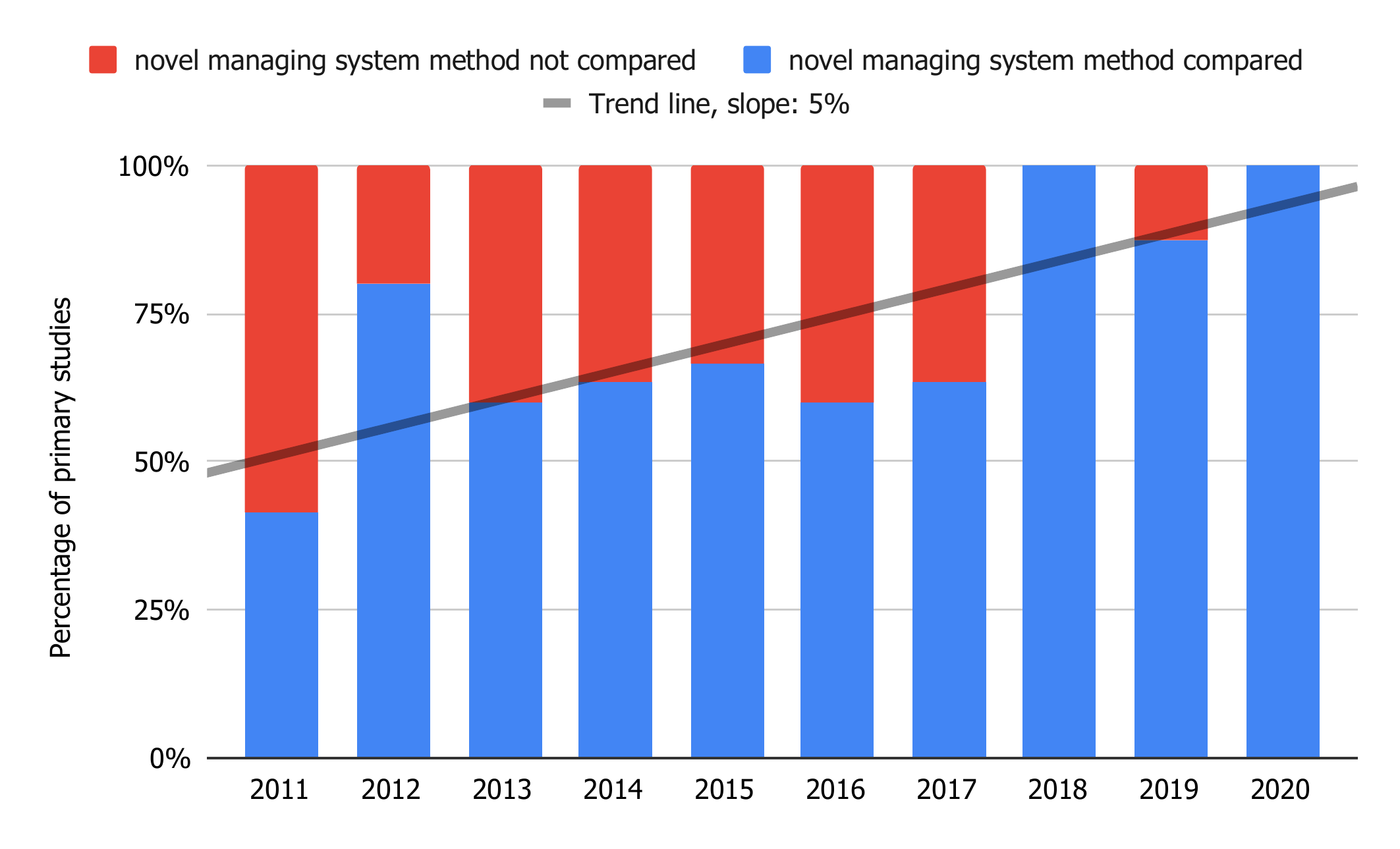}
\caption{Percentage of primary studies per year that do (blue) and do not (red) compare a novel managing system approach with at least one other approach.} 
\label{fig:independent_variables_mm}
\end{figure}

To conclude, we measured the percentage of primary studies that do compare a newly proposed managing system approach with at least one 
other approach, which can for instance be a state-of-the-art approach, primitive adaptation, or the theoretical optimal.
Fig.~\ref{fig:independent_variables_mm} shows the results over the years. The graph clearly illustrates that researchers have increasingly compared new contributions with existing approaches or other baselines. By applying regression on the data, we identified an average yearly increase of 4.67\% (rounded 5\%) over the 10 years (from 51.3\% in 2011 to 93.3\% in 2020). 
This confirms an increasing maturity in the evaluations of new contributions in self-adaptive systems presented \mbox{at SEAMS, which has now reached an excellent level.} 

\vspace{5pt}
\begin{tcolorbox}[size=small]
\small 
\textbf{Answer to RQ2: What is the Experimental Design?} Only one out of three studies provides a well-defined formulation of the evaluation problem, mostly using research questions. Experiments use independent variables for all parts of self-adaptive systems, with most factors related to the managing system. The dominant types of dependent variables are time behavior, functional behavior, and resource utilization, typically of managed systems. New contributions are increasingly compared with other approaches. 
\end{tcolorbox}

\subsection{How are Experiments Operated?}

To answer RQ3, we collected data about the managed system and whether an artifact was used (\refDataItem{ManagedSystem}), its nature  (\refDataItem{ManagedSystemNature}), data provenance, \ie sources of data related to users or environment (\refDataItem{DataProvenance}), and representations of uncertainty (\refDataItem{Uncertainty}).  

\begin{figure}[t]
\centering
\includegraphics[width=1\columnwidth]{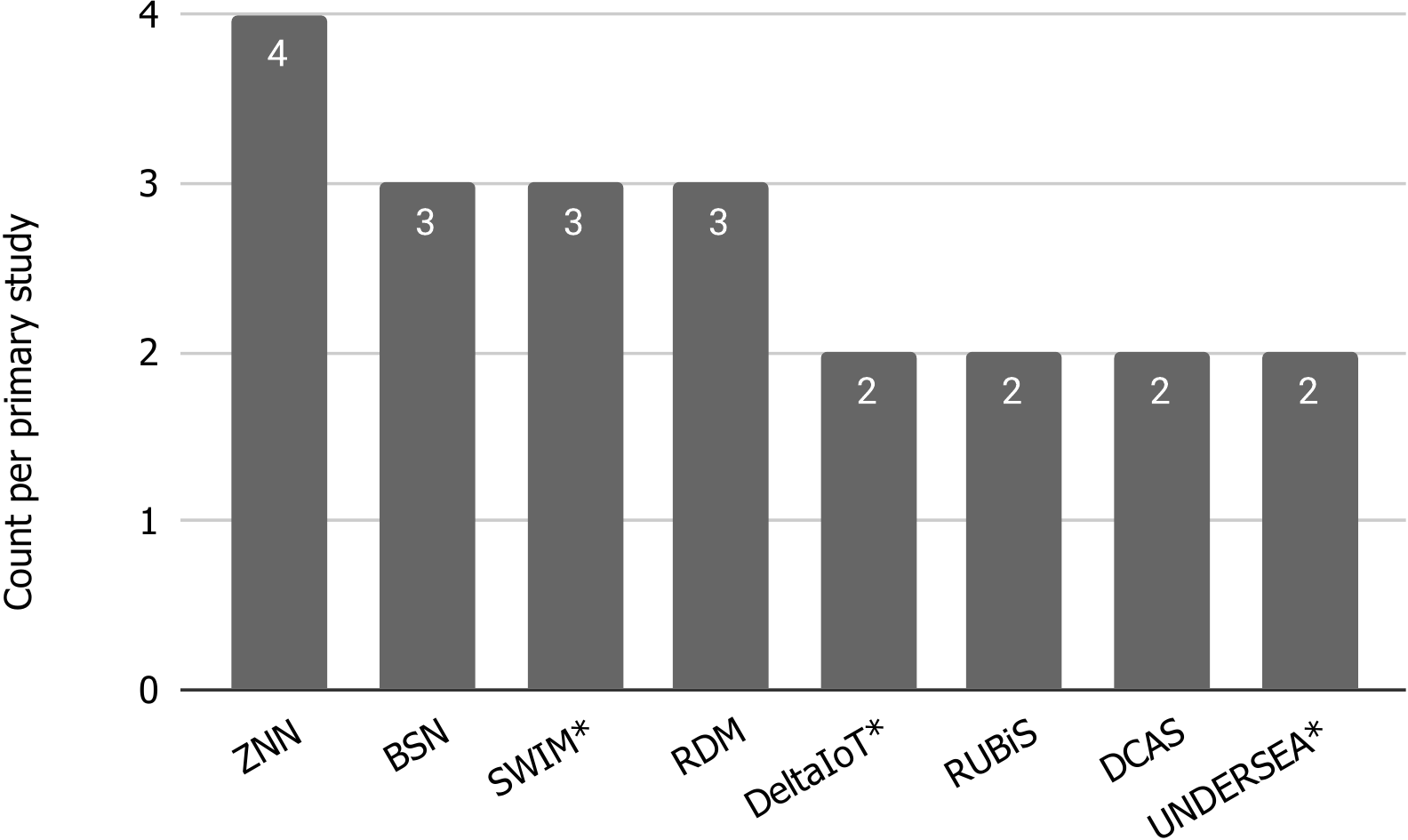}
\caption{Named managed systems (\refDataItem{ManagedSystem}) used in at least two primary studies (43 studies provided no specific name).}
\label{fig:managed_system}
\end{figure}

Fig.~\ref{fig:managed_system} shows the counts of managed systems (\refDataItem{ManagedSystem}) that were used in experiments reported in at least two primary studies. 
The managed systems marked with an asterisk are formally approved SEAMS artifacts. 
Of the total 82 primary studies, 39 (46\%) provided a clearly described name for the managed system. 
The remaining 43 studies provided no specific name for the managed system. 
Not surprisingly, ZNN has been used most frequently, BSN (Body Sensor Network), SWIM, and RDM (Remote Data Mirroring) are each used in three studies, and DeltaIoT, RUBiS (Rice University Bidding System), DCAS (Data Acquisition and Control Service) and UNDERSEA are each used in two studies. 
In total, 10 primary studies used at least one artifact in their evaluation in the period from 2016 to 2020 (\ie 26\% of the primary studies in this period).\footnote{The SEAMS Call for Artifacts was introduced in 2015.} 
This result underpins the usefulness of the SEAMS artifacts in the evaluation of new contributions.

\begin{figure}[t]
\centering
\includegraphics[width=1\columnwidth]{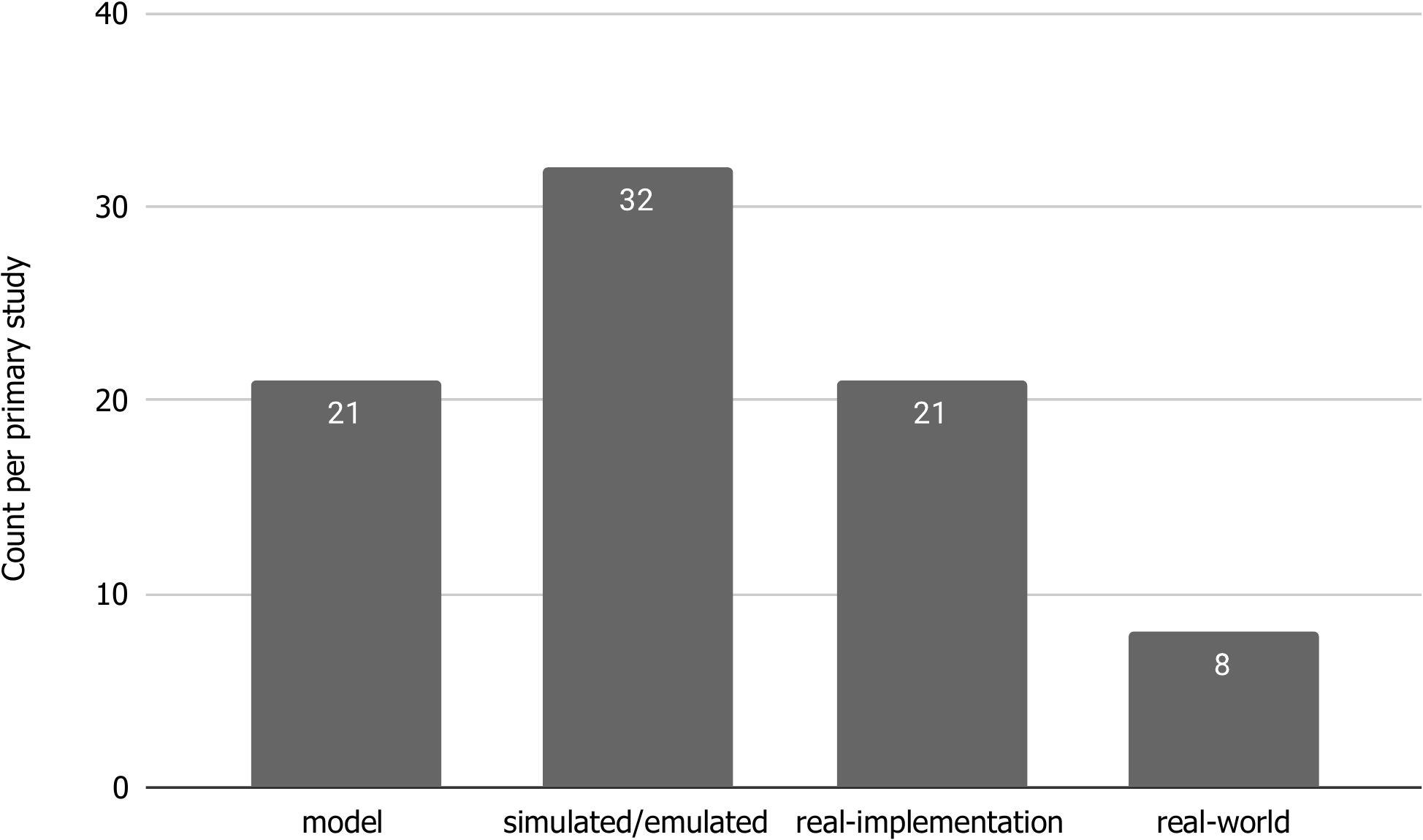}
\caption{Types of managed systems used per study (\refDataItem{ManagedSystemNature}).}
\label{fig:managed_system_nature}
\vspace{-.5em}
\end{figure}

Fig.~\ref{fig:managed_system_nature} shows the  different types of managed systems  (\refDataItem{ManagedSystemNature}) used in the primary studies. 
Thirty-two studies (39\%) used a simulation or emulation of a managed system. 
For example, Gerasimou \etal~\cite{Gerasimou+2014} use UNDERSEA, a simulator of unmanned underwater vehicles. 
Twenty-one studies (38\%) used a model to represent a managed system. 
For example, Incerto \etal~\cite{Incerto+2016} represent a three-tier managed system as a Queuing Network to evaluate performance adaptation. A real implementation of a managed system
was used in 21 studies (26\%). 
This category includes implemented systems based on a model of a real application. 
For example, Barna \etal~\cite{Barna+2017} use LEGIS, a distributed navigation service based on Google maps. 
On the other hand, in eight studies (10\%) the managed system was a real-world application. 
This category refers to systems that have been used in practice with real users (but not necessarily for the experiments). 
An example are open-source mobile apps that are used in~\cite{Riganelli+2017}. 
These results show that a relevant number of experiments rely on real implementations of managed systems, yet with opportunities to further improve on the use of real-world systems in experiments. 

The results for data provenance (\refDataItem{DataProvenance}) show that a majority of 99 experiments (71\%) use synthetic data to represent users or the environment. 
For example, Guerriero \etal~\cite{Guerriero+2018} randomly generate consumer transactions, while Moreno \etal~\cite{Moreno+2017} generate server boot latencies from normal distributions.
Twenty-eight experiments (20\%) use emulated data to represent users or the environment. 
For example, Shin \etal~\cite{Shin+2020} emulate a data traffic profile specified by their industry partner to create load on the managed network.

Notably, only a small fraction of the experiments (13, \ie 9\%) use real-world data to represent users or the environment. 
An example is~\cite{Kistowski+2015} that uses real-world workload traces of the Internet Traffic Archive, Bibsonomy, and Wikipedia to extract load profiles. 
These results show that there is room for improvement to represent users and the environment more realistically in experiments of self-adaptive systems.   

\begin{figure}[t]
\centering
\includegraphics[width=1\columnwidth]{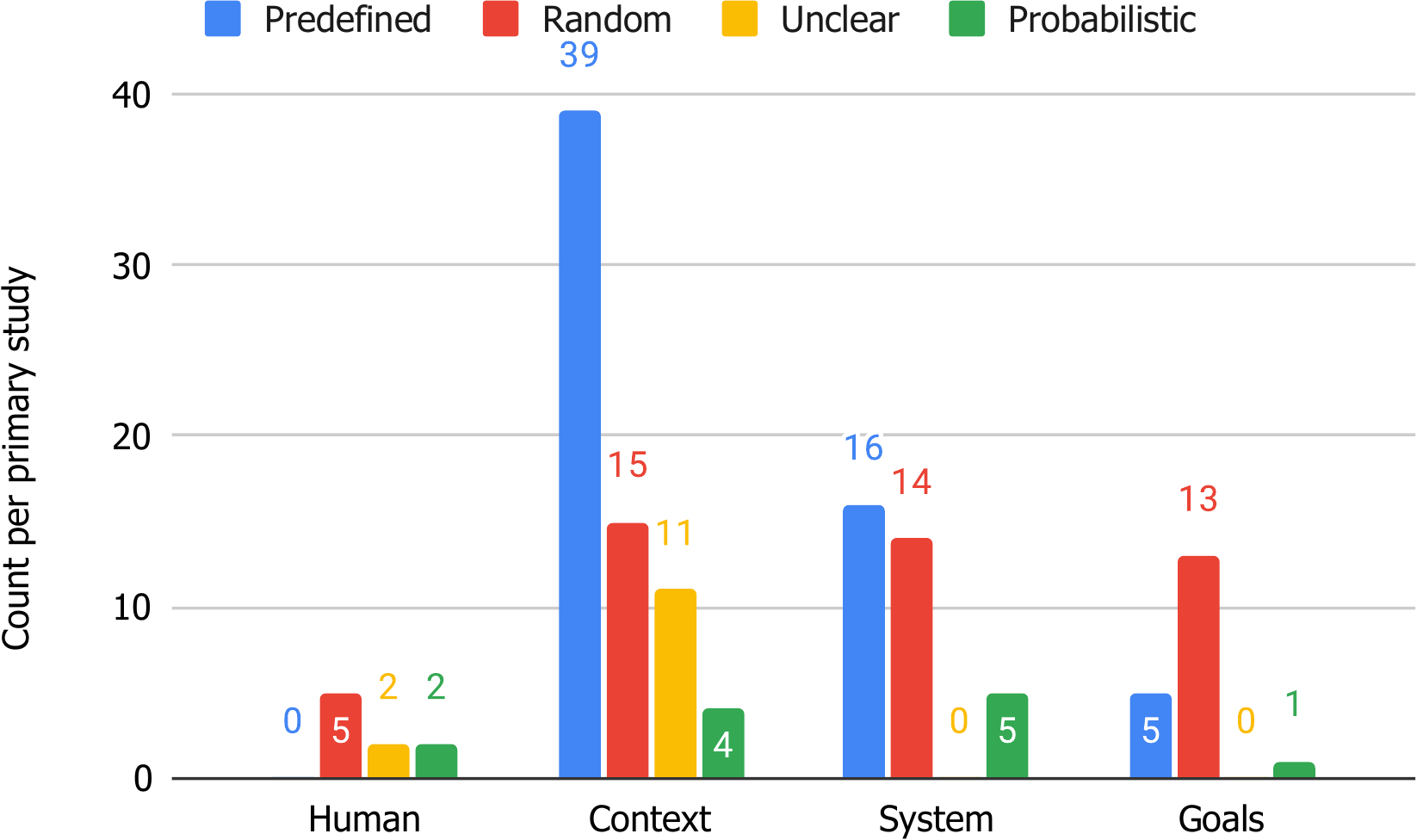}
\caption{Representation of uncertainty per study  (\refDataItem{Uncertainty}).}
\label{fig:uncertainty}
\vspace{-.5em}
\end{figure}

Fig.~\ref{fig:uncertainty} shows the results we obtained for the representation of uncertainty (\refDataItem{Uncertainty}). In total, we extracted 132 representations of uncertainty used in the experiments of the primary studies.\footnote{Note that if two experiments of the same primary study used the same uncertainty representation, this was counted once.} 
We could group these 132 representations in four types. 
The most frequently used type is \textit{uncertainty in the context} that was used 68 times (52\% of all represented uncertainties).
As an example, Jamshidi~\etal~\cite{Jamshidi2019} consider the uncertainty of having obstacles appearing in the robot's environment. 
\textit{Uncertainty in the system} was used 35 times (27\%).\footnote{Uncertainty of the system relates to the managed system, either represented in the managed system itself or in a runtime model of the managed system used by the managing system.} 
For example, Incerto~\etal~\cite{Incerto+2016} address uncertainty of the system in terms of random faults of servers and network links.
Only a few studies \textit{considered uncertainty of goals} (20 studies, 15\%) and humans (9 studies, 7\%). 
For example, Camara~\etal~\cite{Camara+2020} randomly assign missions to robots, while Tun~\etal~\cite{Tun+2018} randomly select invitees (users) for sharing files. 

Uncertainty in the experiments is mostly represented by predefined values (46\% of all uncertainty representations).
Such values can be based on common properties or characteristics of the represented topic. 
For example, Zanardi and Capra~\cite{Zanardi+2011} consider different predefined adaptation thresholds to model the uncertainty related to goals.
Uncertainties (35\%) were represented by randomly selected values 64 times (\eg \cite{Tun+2018}). 
Only 12 uncertainties (9\%) were represented by probabilistically selected values (\eg \cite{Moreno+2017,Kinneer2018}).
   
The results of \refDataItem{Uncertainty} show that uncertainty is commonly considered in experiments of self-adaptive systems. Yet, the emphasis is on uncertainty in the context and  system. There is room for improvement by putting more emphasis of representing uncertainties related to goals and humans in experiments. 
 
\vspace{5pt}
\begin{tcolorbox}[size=small]
\small 
\textbf{Answer to RQ3: How are Experiments Operated?} Artifacts are increasingly used in experiments of self-adaptive systems. The managed system is mostly simulated or emulated. Yet, one on three studies uses system implementations, but the number of real-world systems or prototypes of such systems remains relatively low. Most data of users and the environment used in experiments is synthetically generated. Studies commonly consider uncertainties in the context and the system, mostly represented by selected values (rather than randomly or probabilistically). Uncertainties related to goals and humans are not frequently considered. 
\end{tcolorbox}

\subsection{How is the Experiment Data Analyzed?}

To answer RQ4, we collected data about the types of analysis applied (\refDataItem{AnalysisType}), the answers provided for the evaluation problem (\refDataItem{AnswerToEvaluationProblem}) and the discussion of treats to validity (\refDataItem{ValidityThreats}).  

The results for the type of analysis (\refDataItem{AnalysisType}) performed in the experiments show that a majority of 62 studies (44\%) used some form of exposition or narrative to analyze and discuss the experiment results. 
For example, Weisenburger~\etal~\cite{Weisenburger+2017} plot the timeseries of latency and bandwidth obtained by their approach and the baseline under different settings and discuss the observed behavior.
Fifty-nine studies (42\%) used descriptive statistics to show or summarize data in a meaningful way (\eg using tables, graphs and charts), which allows identifying patterns that might emerge from the data. 
As an example, Sousa~\etal~\cite{Sousa+2017} analyze their experiment by calculating statistics (average, standard deviation, maximum, minimum) of execution times over 12 runs. 
Finally, 19 studies (14\%) used statistical tests to analyze the data of the experiment and draw conclusions. 
A statistical test provides a systematic mechanism for making a quantitative decision about the outcome of an experiment; for instance to determine whether there is enough evidence to reject a null hypothesis. 
For example, Fredericks~\etal\cite{Fredericks+2014} define a hypothesis to test statistically whether a difference exists between the result of their proposed approach and a baseline. 
We also extracted data about whether the choice for a statistical test was motivated and found that only 12 of the 19 studies do so. 
The results show that a substantial part of the studies still apply informal approaches for the analysis of data of experiments.
There is room for applying more rigorous methods of analysis of data obtained from experiments in self-adaptive systems.  

Of the 28 studies that formulated evaluation problems using either research questions or hypothesis, only 19 (68\%) provided an explicit answer to the evaluation problem. 
For example, Chen~\cite{Chen2019} summarizes the key findings of each of the specified four research questions, while Fredericks~\etal\cite{Fredericks+2014} explain the rejection of the specified null hypothesis based on a statistical test. 
The numbers underpin a need for improvement on reporting research findings from experiments, in particular providing answers to the research questions under study. 

\begin{figure}[t]
\centering
\includegraphics[width=1\columnwidth]{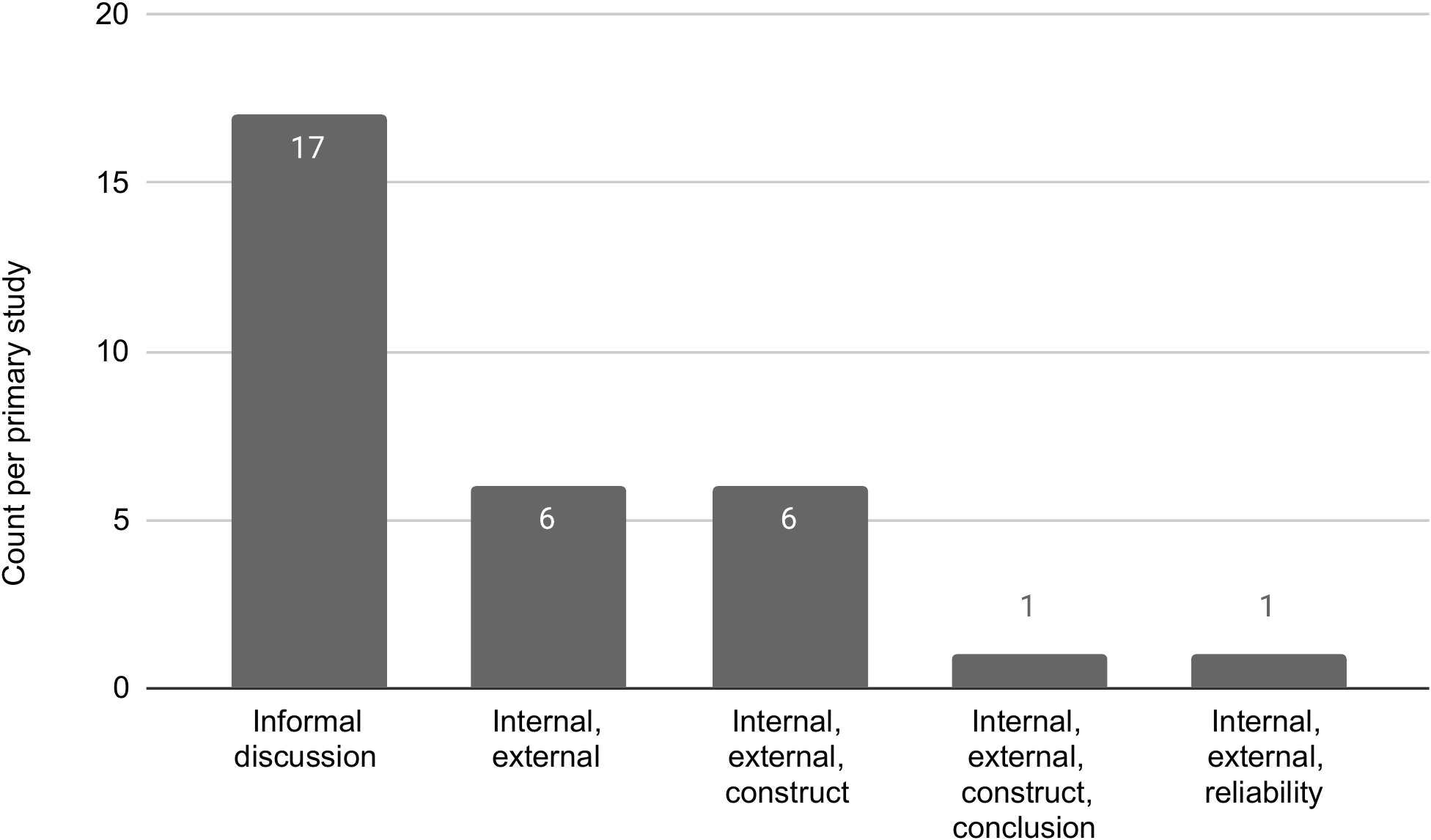}
\caption{Discussion of threats to validity per study (\refDataItem{ValidityThreats}).}
\label{fig:threats_to_validity}
\end{figure}

Fig.~\ref{fig:threats_to_validity} shows the results for whether and how threats to validity of experiments (\refDataItem{ValidityThreats}) are discussed in primary studies. 
In total only 31 primary studies (38\% of all primary studies) provided some discussion of validity threats.
Seventeen of these studies (55\% of the 31 studies that discuss validity threats) provide an informal discussion of validity threats without referring to any particular types of threats. 
As an example, Sousa~\etal\cite{Sousa+2017} discuss the limitations of their experiments in a separate section in an informal way. 
The most reported validity threats are internal and external validity, both discussed in 14 studies (45\% of the 31 studies that discuss validity threats). 
For example, Jamshidi \etal~\cite{Jamshidi2017} discuss internal and external validity and their attempt to mitigate these threats, and discuss remaining limitations. 
Of these 14 studies, seven also discuss construct validity. 
As an illustration, Kumar~\etal\cite{Kumar+2020} mention construct validity threats related to the employed metric and evaluation methods, and how they mitigate them.
Only one study~\cite{Quin2019} discusses reliability pointing out that reproducing the results of the study may be affected by randomness included in the simulation setup. 
Acknowledging and discussing validity threats is key for future research as they pinpoint potential issues with the experimental design and the causal relationships and generalization of results. 
Hence, there is room for improvement on discussing validity threats for experiments of self-adaptive systems.

\vspace{5pt}
\begin{tcolorbox}[size=small]
\small 
\textbf{Answer to RQ4: How is the Experiment Data Analyzed?} A small half of the primary studies use an informal approach to analyze the data obtained from experiments. Another small half uses descriptive statistics and only a fraction of the studies uses statistical tests. Only a limited number of the primary studies provide explicit answers to the research problems they tackle. Explicitly discussing threats to validity is not common practice in experimental research of self-adaptive systems.  
\end{tcolorbox}

\subsection{How are Experiment Results Packaged?}
To answer RQ5, we collected data about the availability of the experimental results (\refDataItem{ResultsAvailable}), and the degree of reproducibility (\refDataItem{CodeAvailable}) of experiments.  

Only 11 of the primary studies (13\%) made the evaluation results of their experiments publicly available (\refDataItem{ResultsAvailable}). 
Examples of studies that provide evaluation results are~\cite{Incerto+2016,Shin+2020}, where results are made available with a replication package.
Making experiment data public allows for verifying findings and experimental reuse, and lowers the barriers to meta-studies.\footnote{www.elsevier.com/connect/should-research-data-be-publicly-available}
Hence, there is room for improvement here, but this is a general problem and applies also to other research fields than self-adaptive systems (\eg~\cite{Pimentel+2019}). 

The results we obtained for the degree of reproducibility (\refDataItem{CodeAvailable}) of experiments reported at SEAMS echo those of \refDataItem{ResultsAvailable}. Only nine studies (11\%) provide a full replication package, while 14 other studies (17\%) provide the code used in experiments. 
Although it is a foundation of science, replication is a recurrent issue in empirical software engineering in general and this applies also to software engineering of self-adaptive systems. 
The results of this study show a slight improvement in terms of reproducibility compared to the results of the earlier study~\cite{weyns2012claims} that looked at research presented at SEAMS before 2012. 
There, 14\% of the studies provided partial material for repeatability and only 2\% provide a full replication package.

All in all, there remains substantial room for further improvement on providing replication packages facilitating cross-validation and comparison across studies.

\vspace{5pt}
\begin{tcolorbox}[size=small]
\small 
\textbf{Answer to RQ5: How are the Experiment Results Packaged?} Only a small fraction of the primary studies make the data of their experiment publicly available. Similarly, the degree of reproducibility remains low, only one study on ten offers a full replication package to the community. 
\end{tcolorbox}

\section{Discussion}\label{sec:insights}

We start this section with summarizing insights on the specifics of experiments in self-adaptive systems. Then we offer
suggestions to improve future experiments in self-adaptive systems. Finally, we discuss threats to validity of this study.

\subsection{Experiments in Self-adaptive Systems vs Other Systems}

\begin{figure}[h]
\centering
\includegraphics[width=1\columnwidth]{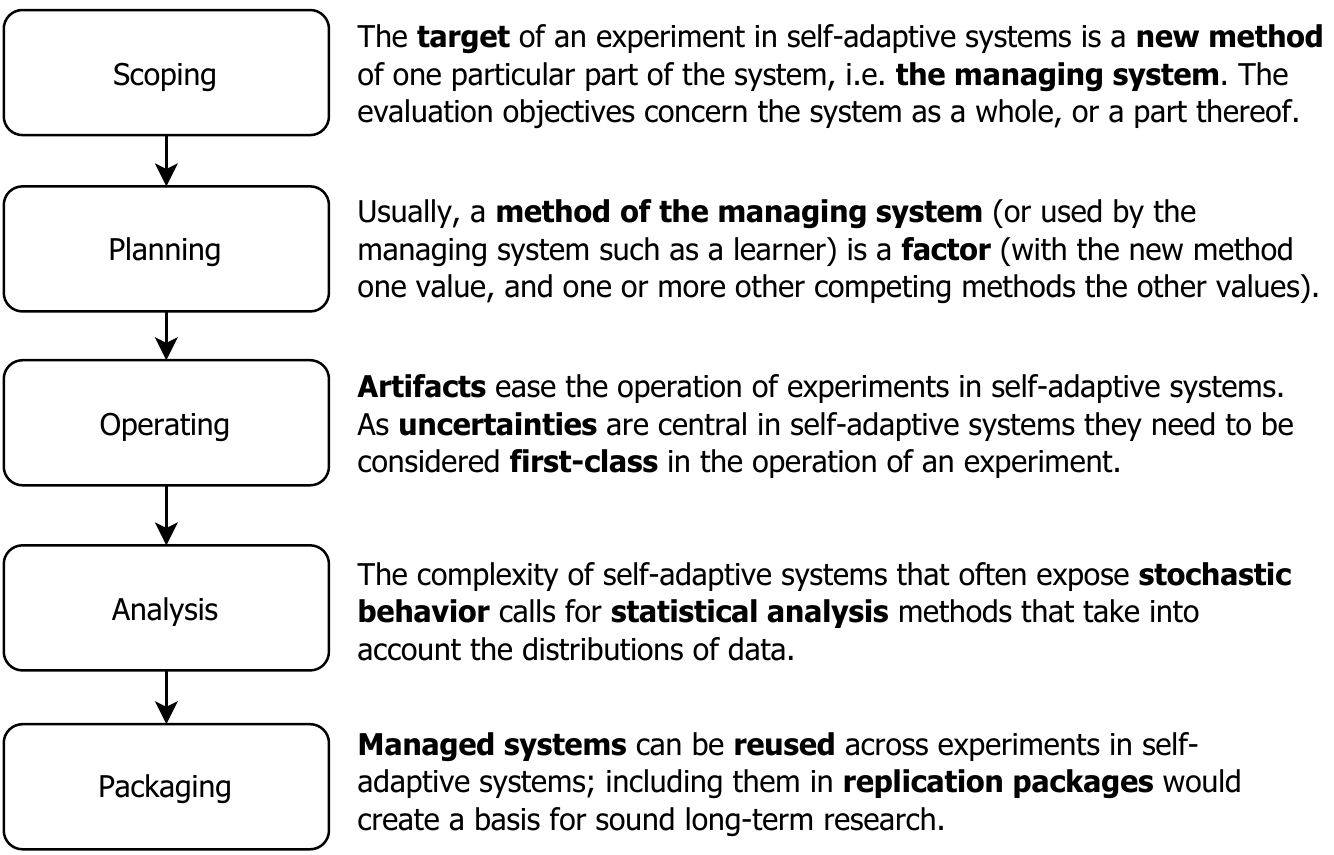}
\caption{Main insights on the process of experiments.}
\label{fig:insights}
\vspace{-1em}
\end{figure}

This mapping study provides a number of insights on the specifics of experiments in self-adaptive systems compared to other systems. In Fig.~\ref{fig:insights}, we list these insights based on the five steps of the process of an  experiment~\cite{Wohlin+2012}. 

\subsection{Suggestions for Improving Future Experiments} 

In what follows, we offer a number of suggestions that we obtained from this study as impetus to improve future experiments in self-adaptive systems. Experiments on new methods for the monitor and execute stages of MAPE-K feedback loops require attention. 
Security, privacy, humans in the loop, and ethical considerations are widely regarded as key concerns of modern software systems but they are not well studied in our field, hence they deserve attention. 
We can improve on better formulating the evaluation problems we tackle. We often use different types of design compared to experiments in traditional software systems; this may either point to a lack of maturity or the need for designs specific to experimentation in self-adaptive systems; this deserves further investigation. We observe an increasing trend in the use of real-world managed systems; experiments would benefit from pushing this trend further. Uncertainty is a complex phenomenon but essential to self-adaptation; there are plenty of opportunities to enhance on how we represent uncertainties in experiments. There is substantial room to improve on the analysis of experimental data; in particular by applying statistical techniques. Experiments in self-adaptive systems would benefit from a more rigorous description of different types of validity threats, and making replication packages available for the community. We hope that these empirically grounded suggestions will help the community to improve the way we evaluate self-adaptive systems in the future.  

\subsection{Threats to Validity} 

While following a systematic approach based on a protocol, this study has some possible threats to validity. 

\textit{Internal validity} refers to the extent to which a causal conclusion can be made based on the study. Determining whether a paper contained an experiment was sometimes not easy as some information may be implicit. In addition, the extraction of detailed information about experiments, in particular identifying different types of variables was also not always easy. To mitigate this threat, we took the following measures. (i) All papers where checked for inclusion or exclusion by at least two researchers. (ii) The primary studies were split in three parts; for the first part (10\% of the studies) data was extracted in parallel by two researchers and decisions were made based on consensus; in case of conflicts a third researcher was consulted to make a decision; for the two other parts, data was extracted by one researcher and crosschecked by another (and if needed by a third). (iii) Data analysis was jointly done by all researchers in collaboration.

\textit{External validity} refers to the extent to which the findings can be generalized. By considering only full papers presented at SEAMS, we obviously can only draw conclusions for this venue. However, as argued in the summary of the protocol (Section~\ref{sec:protocol}), the papers presented at SEAMS provide a representative sample of software engineering research of self-adaptive systems. Furthermore, the draft of the ACM SIGSOFT Empirical Standards consider focusing on a single venue an acceptable deviation of performing secondary studies~\cite{ralph2020acm}. 
Nevertheless, to strengthen the validity of our study, \mbox{a broader search for primary studies would enhance validity.}

\textit{Construct validity} refers to the extent to which we obtained the right measure and whether we defined the right scope in relation to the topic of our study. There is threat that the reporting of experiments in some papers may not be of sufficient quality. However, since SEAMS became a symposium in 2011, we believe that papers that were accepted as full papers provide a sufficiently good quality of reporting. We acknowledge that an additional quality check based on the quality criteria for reporting studies as \eg used in~\cite{DYBA2008833,7929422} may help improving the validity of our results.  

\textit{Reliability} refers to the extent to which we can ensure that our results are the same if our study would be conducted again. An obvious threat is a potential bias of the researchers involved in our study, in particular when collecting and analyzing data of primary studies. To address this threat, we used a protocol that we carefully followed. In addition, we have made all the material of the study available for other researchers. 

\section{Conclusions}\label{sec:conclusions}

This mapping study aimed at answering the question ``How do we evaluate self-adaptive software systems?'' with a focus on technology-oriented experiments presented at SEAMS from 2011 till 2020. Results show that experiments in self-adaptive systems do follow standard practice on empirical research in software engineering, but at the same time also have some specifics that deserve special attention across the stages of the experimental process. These specifics are essentially based on characteristics of self-adaptive systems, such as the evaluation target that is associated with the managing system, and the presence of uncertainties that require attention in experiment design and analysis. 
Our study allowed us to provide a number of suggestions for improving experimental evaluations of self-adaptive systems. We hope that these suggestions and the results obtained from our study trigger reflection in the community on doing future experiments with even more maturity.

\bibliography{ref}
\bibliographystyle{plain}

\end{document}